\documentclass[useAMS,usenatbib]{mn2e}

   \usepackage{graphicx}
   \usepackage{amsmath}
   \usepackage{color}

\title[A search for H$_{\rm 2}$O masers in nearby QSOs]{A search for H$_{\rm 2}$O maser emission in nearby low-luminosity QSO host galaxies}

\author[S. K\"onig, A. Eckart, et al.]{S. K\"onig$^{1}$\thanks{E-mail:
skoenig@dark-cosmology.dk}\thanks{Current address: Dark Cosmology Centre, Juliane Maries Vej 30, DK-2100 Copenhagen, 
Denmark}, A. Eckart$^{1,2}$, C. Henkel$^{2}$, M. Garc\'{i}a-Mar\'{i}n$^{1}$\\
$^{1}$I. Physikalisches Institut, Universit\"at zu K\"oln, Z\"ulpicher Strasse 77, D-50937 K\"oln, Germany\\
$^{2}$Max-Planck-Institut f\"ur Radioastronomie, Auf dem H\"ugel 69, D-53121 Bonn, Germany}
\begin{document}

%\date{Accepted 1988 December 15. Received 1988 December 14; in original form 1988 October 11}

\pagerange{\pageref{firstpage}--\pageref{lastpage}} \pubyear{2002}

\maketitle

\label{firstpage}

\begin{abstract}
A sample of nearby galaxies hosting low-luminosity type 1 quasi\,-\,stellar objects (QSOs), previously studied in CO and H{\sc{I}} emission lines, has been searched for maser emission. Using the Effelsberg 100-m telescope, we observed 17 sources with redshifts of $z$\,$\leq$\,0.06 and $\delta$\,$>$\,-\,30\degr\ for emission in the 22~GHz water vapor maser transition. The sample objects have been drawn from a wide-angle survey for optically bright QSOs (Hamburg\,/\,ESO survey, HES). No host galaxies from the sample have been detected in the water maser emission line. In this paper we review the discussion on the reasons for H$_{\rm 2}$O megamasers being rarely found in Seyfert~1 galaxies. Eight of them are bulge dominated and probably of elliptical type (E/S0), whereas 6 have spiral geometry. Three of the objects seem to be in a phase of merging/interaction. We found 3\,$\sigma$ upper limits for the flux density of 27 to 60~mJy at spectral resolutions of $\sim$\,0.43~km\,s$^{\rm -1}$. We furthermore find that the viewing angle to the line of sight to the galaxy, under which the probability to detect megamaser emission is highest, is about 6\degr.
\end{abstract}

\begin{keywords}
masers -- galaxies:active -- quasars:emission lines -- radio lines:galaxies
\end{keywords}

\section{Introduction}

22~GHz ($\lambda$ $\sim$ 1.3~cm) water vapor masers are excellent tracers of physical conditions, such as temperature ($T_{\rm kin}$\,$>$\,400~K) and density
($n(H_{\rm 2}$)\,$>$\,10$^{\rm 7}$~cm$^{\rm -3}$) of the molecular gas, in the highly obscured innermost parts of active galactic nuclei. Extragalactic water masers with isotropic luminosities $L_{\rm H_2O}$\,$>$\,10~$L_{\sun}$ are classified as
\textit{megamasers}. But since the apparent luminosity is derived under the assumption of isotropic emission, the true luminosity may be smaller by several orders of magnitude \citep[see e.g., the recent review by][]{lo05}. \citet{hen05} found a transition near $L_{\rm H_2O}$\,=\,10~$L_{\sun}$ between weaker masers mostly related to star formation and stronger masers associated with active galactic nuclei. Furthermore, they found a correlation between the luminosity in the infrared and the total isotropic water maser luminosity.\\
\indent
Towards Seyfert and LINER galaxies, H$_{\rm 2}$O megamasers are used to probe the small scale structure and kinematics of accretion disks or tori, to obtain the masses of the central engines through a determination of Keplerian rotation \citep{kuo11}, and to derive geometric distances to the parent galaxies \citep[e.g.,][]{miyo95,her99,bra10}. Complementing these so-called \textit{disk-masers} are the \textit{jet-masers} \citep[e.g.,][]{clau98,peck03}, which are arising from the interaction of the nuclear jet with a cloud being accidentally located along the line-of-sight to the jet. In one source, Circinus, a nuclear \textit{outflow-maser} \citep{green03b} was also detected. To date, H$_{\rm 2}$O vapor megamasers are found in 10\% of observed AGNs in the local universe \citep{bra04}. They are almost exclusively found in Seyfert 2 and LINER type galaxies \citep{bra97,kon06a}, i.e. mostly in spirals. This can be interpreted in terms of the unified scheme in that way that AGN activity is required for megamaser emission and that the orientation of the nuclear disk plays a role. The nuclear disks of Seyfert 2 galaxies, for example, are seen roughly edge-on. Because the nuclear disks of spirals are not aligned with the large scale morphology of the parent galaxy \citep{ulve84}, a convincing correlation between the inclination of the large scale disks and the presence of megamasers is not apparent.\\
\indent
The first quasar discovered to emit water megamaser emission at 22~GHz, \citep[J080430.99+360718.1,][]{bar05}, was classified as a type 2 quasar. The fact that the second H$_{\rm 2}$O megamaser emitting quasar is a type 1 quasar \citep[MG~J0414+0534,][]{imp08}
brings the (statistically non-significant) detection ratio for quasars to 1:1. Since QSOs have more massive cores than Seyfert galaxies, the chance to find maser emission in type 1 or type 2 quasars could totally differ from what is known for Seyfert galaxies. So the question is: Could the detection by \citet{imp08} just be a serendipitous one, or could this detection of water megamaser emission in a type 1 quasar be trend-setting for the local universe? In order to find an answer to this question, the sample at hand was searched for water masers in southernly nearby low-luminosity type 1 QSO host galaxies. In
Sect.\,\ref{section:maser_sample} the sample is described, Sect.\,\ref{section:maser_obs} is devoted to the characterization of the observations and Sect.\,\ref{section:maser_results} comprises the results and discussion. Sect.\,\ref{section:maser_conclusions} gives a short summary.\\
\indent
Unless otherwise stated, H$_{\rm 0}$\,=\,75~km\,s$^{\rm -1}$\,Mpc$^{\rm -1}$ and q$_{\rm 0}$\,=\,0.5 are assumed throughout the paper.

%__________________________________________________________________

\section{The sample} \label{section:maser_sample}

We observed sources from a volume limited sample of nearby low-luminosity QSO host galaxies taken from the Hamburg\,/\,ESO survey \citep[HES;][]{wis00}. The HES is a wide angle survey for optically bright QSOs, with a well-defined flux limit of $B$$_{\rm J}$\,$<$\,17.3, varying
from field to field, and a redshift coverage of 0\,$<$\,$z$\,$<$\,3.2. The upper redshift limit was set to z\,$<$\,0.06 to ensure the observability of the CO(2-0) rotation vibrational band head absorption line in the near infrared K-band. \\
\indent
It is important to note that no luminosity discrimination between QSOs and Seyfert 1 galaxies was applied by the HES. This is important 
for the absolute brightness distribution of the sample here referred to as ``nearby low-luminosity QSO sample''. Our sample clearly probes the low 
luminosity tail of the local quasar luminosity function \citep{koehler97}. All objects in the sample have absolute $B$$_{\rm J}$ magnitudes 
exceeding the traditional boundary $M$$_{\rm B}$ $\sim$ -22 between higher luminosity QSOs and lower luminosity Seyfert 1 galaxies. To respect the 
commonly used definition of the term ``QSO'', we explicitely use the term ``low-luminosity QSO'' throughout the paper for objects identified in QSO 
surveys that may be fainter than the traditional boundary magnitude. Not only the HES but also the PG Bright Quasar Survey provide low-luminosity 
QSOs in their samples.\\
\indent
39 of the altogether 99 sources, with a declination $\delta$\,$>$\,-\,30\degr, have been searched for CO emission with the IRAM~30\,m telescope on the Pico Veleta (Spain) and the SEST (Swedish ESO-Submillimeter Telescope) in La Silla (Chile) \citep{bert07}. A follow-up study of the 27 CO detected sources was carried out in the H{\sc{I}} 21~cm line with the Effelsberg 100-m telescope. About 45\% of the sources observed in H{\sc{I}} show emission at this frequency \citep{koenig09}. A more detailed description of the 'nearby QSO sample' can be found in \citet{bert07}.\\
\indent
In this paper we study the 17 sources of the original CO subsample which are most luminous at infrared wavelengths (Table\,\ref{tab:properties}), with recession velocities ranging from 7900~km\,s$^{-1}$ to 18\,200~km\,s$^{-1}$, in the 22~GHZ ($\lambda$ $\sim$ 1.3~cm) H$_{\rm 2}$O maser transition. Morphological classifications for our sample sources range from elliptical to spiral galaxies, galaxies that appear to be involved in merger activity and a large fraction of our sample galaxies has companions. Jets are not known for any of the sample galaxies. In addition to the 17 nearby QSO hosts we observed W3(OH) and  Orion$-$KL as ``control sources'' in the 22.235~GHz H$_{\rm 2}$O maser emission line.

\begin{table*}%[!h]
\begin{minipage}[h]{\textwidth}
\centering
\renewcommand{\footnoterule}{}
\caption{\small
 Sources searched for H$_{\rm 2}$O maser emission.}
\label{tab:properties}
\tabcolsep0.1cm
\begin{tabular}{lcccrcccrrccc}
%\begin{longtable}{lcccr}
\noalign{\smallskip}
\hline
\noalign{\smallskip}
\hline
\noalign{\smallskip}
%---------------------------------------------------------------------------
Object &  & RA(2000) & DEC(2000) & v$_{\rm 0}$\,(LSR) & z & t$_{\rm obs, on}$\footnote{total observation time
spent on the source} & rms & Intensity I\footnote{Upper limits and errors of the intensity represent 1\,$\sigma$ values.} & log 
L$_{\rm H_2O}$\footnote{$L$$_{\rm H_2O}$/[$L$$_{\sun}$]\,=\,0.023\,$\times$\,$\int$$S$d$V$/[Jy\,km\,s$^{\rm -1}$]\,$\times$\,$D$$_{\rm L}^{\rm 2}$/[Mpc$^{\rm 2}$], \citep{hen05}} & AGN\footnote{taken from the NED.} & Morphological\footnote{E denotes elliptical morphology; S represents a spiral morphology; M stands for possible mergers or merger remnants; R denotes ringed objects; C marks galaxies with other extragalactic sources within a projected distance of up to 340~kpc}& M$_{\rm BH}$\footnote{References: for HE\,0040$-$1105, HE\,0114$-$0015: \citet{green06a}; for HE\,0212$-$0059: \citet{green06b}; for HE\,0232$-$0900, HE\,2302$-$0857: \citet{oneill05}; for HE\,1011$-$0403: \citet{wang01}}\\
\noalign{\smallskip}
        & &  [h] [m] [s] & [\degr] [\arcmin] [\arcsec] & [km\,s$^{-1}$] & & [min] & [Jy] & [Jy\,km\,s$^{\rm -1}$] & [L$_{\sun}$] & 
        Type & Type & [M$_{\sun}$]\\
%---------------------------------------------------------------------------
\noalign{\smallskip}
\hline
\noalign{\smallskip}
HE\,0021--1819 & & 00:23:55.3 & --18:02:50 & 15\,954 & 0.053 & 27 & 0.013 & $<$ 0.53\,$\pm$\,0.16 & $<$ 2.75 & Sy~1   & E, C          & -- \\
HE\,0040--1105 & & 00:42:36.8 & --10:49:21 & 12\,578 & 0.042 & 27 & 0.013 & $<$ 0.29\,$\pm$\,0.16 & $<$ 2.28 & Sy~1.5 & E, C          & 10$^{\rm 6.70}$ \\
HE\,0114--0015 & & 01:17:03.6 &  +00:00:27 & 13\,682 & 0.046 & 27 & 0.011 & $<$ 0.02\,$\pm$\,0.14 & $<$ 1.15 & NLSy~1 & E, C, poss. M & 10$^{\rm 6.80}$ \\
HE\,0119--0118 & & 01:21:59.8 & --01:02:25 & 16\,412 & 0.055 & 27 & 0.010 & $<$ 0.39\,$\pm$\,0.13 & $<$ 2.64 & Sy~1.5 & E             & -- \\
HE\,0150--0344 & & 01:53:01.4 & --03:29:24 & 14\,329 & 0.048 & 27 & 0.011 & $<$ 0.39\,$\pm$\,0.14 & $<$ 2.52 &        & M             & -- \\
HE\,0212--0059 & & 02:14:33.6 & --00:46:00 & 7\,921  & 0.026 & 27 & 0.014 & $<$ 0.10\,$\pm$\,0.18 & $<$ 1.43 & Sy~1.2 & E, C          & 10$^{\rm 7.20}$ \\
HE\,0224--2834 & & 02:26:25.7 & --28:20:59 & 18\,150 & 0.060 & 27 & 0.020 & $<$ 0.40\,$\pm$\,0.25 & $<$ 2.75 & Sy~1   & M, C    & -- \\
HE\,0232--0900 & & 02:34:37.7 & --08:47:16 & 12\,886 & 0.043 & 27 & 0.015 & $<$ 0.24\,$\pm$\,0.19 & $<$ 2.21 & Sy~1   & R, M, C       & 10$^{\rm 8.05}$ \\
HE\,0345+0056  & & 03:47:40.2 & +01:05:14  & 8\,994  & 0.031 & 27 & 0.016 & $<$ 0.37\,$\pm$\,0.20 & $<$ 2.12 & Sy~1   & E, C          & -- \\
HE\,0433--1028 & & 04:36:22.2 & --10:22:33 & 10\,651 & 0.036 & 54 & 0.011 & $<$ 0.49\,$\pm$\,0.14 & $<$ 2.36 & Sy~1   & S             & -- \\
HE\,0853--0126 & & 08:56:17.8 & --01:38:07 & 17\,899 & 0.060 & 54 & 0.008 & $<$ 0.76\,$\pm$\,0.10 & $<$ 3.01 & Sy~1   & E, C          & -- \\
HE\,1011--0403 & & 10:14:20.6 & --04:18:41 & 17\,572 & 0.059 & 27 & 0.011 & $<$ 0.28\,$\pm$\,0.14 & $<$ 2.56 & Sy~1   & S, C          & 10$^{\rm 7.03}$ \\
HE\,1017--0305 & & 10:19:32.9 & --03:20:15 & 14\,737 & 0.049 & 93 & 0.007 & $<$ 0.06\,$\pm$\,0.09 & $<$ 1.71 & Sy~1   & S, C, poss. M & -- \\
HE\,1107--0813 & & 11:09:48.5 & --08:30:15 & 17\,481 & 0.058 & 27 & 0.012 & $<$ 1.02\,$\pm$\,0.15 & $<$ 3.12 & Sy~1   & E           & -- \\
HE\,1126--0407 & & 11:29:16.6 & --04:24:08 & 18\,006 & 0.060 & 27 & 0.012 & $<$ 0.10\,$\pm$\,0.15 & $<$ 2.14 & Sy~1   & S, C          & -- \\
HE\,2233+0124  & & 22:35:41.9 & +01:39:33  & 16\,913 & 0.056 & 45 & 0.009 & $<$ 0.20\,$\pm$\,0.11 & $<$ 2.37 & Sy~1   & S, C          & -- \\
HE\,2302--0857 & & 23:04:43.4 & --08:41:09 & 14\,120 & 0.047 & 30 & 0.012 & $<$ 0.12\,$\pm$\,0.15 & $<$ 2.00 & Sy~1.5 & S, C          & 10$^{\rm 8.54}$ \\
W3(OH)         & & 02:27:04.1 & +61:52:22  & --46.4\footnote{Velocity taken from \citet{bro96}.} & 0.000 & 12 & 0.087 & 9\,944\,$\pm$\,0.59 & 1.94 & -- & -- & -- \\
Orion--KL      & & 05:35:14.2 & --05:22:22 & 7.74\footnote{Velocity taken from \citet{mat00}.}   & 0.000 & 18 & 0.040 & 19\,203\,$\pm$\,0.32 & 0.67 & -- & -- & -- \\
resampled\footnote{All maser spectra of the nearby QSO sample sources were resampled and averaged to obtain these values.}               & &     --     &     --       & --    &  --  & -- & 0.001 & $<$ 0.07\,$\pm$\,0.05 & -- &  --    & -- & -- \\
\noalign{\smallskip} \hline
%--------------------------------------------------------------------------
%\end{longtable}
\end{tabular}
\end{minipage}
\end{table*}

%__________________________________________________________________

\section{Observations and data reduction} \label{section:maser_obs}

The observations were carried out in the 6$_{\rm 16}$-5$_{\rm 23}$ 22.235~GHz H$_{\rm 2}$O maser transition on 27 and 28 November 2007 using the Effelsberg 100-m telescope. We used the 18\,-\,26~GHz two\,-\,channel K-band HEMT facility receiver as the frontend in conjunction with a 8192-Channel-Autocorrelator (AK\,90) and a Fast Fourier Transform Spectrometer (FFTS) as backends. The latter provided a bandwidth of 500~MHz with a velocity resolution of $\sim$\,0.8~km\,s$^{-1}$ (2 channels), thus covering a velocity range of approximately 7150~km\,s$^{-1}$. The AK90 in NSPLIT mode 25 consisted of eight channel backends with a somewhat coarser velocity resolution of $\sim$\,1~km\,s$^{-1}$ and an individual bandwidth of 40~MHz. Some of the eight backends were shifted in frequency to cover altogether a total bandwidth of 80~MHz. We used the FFTS in the load switching mode, employing the rotating horn of the 1.3~cm primary focus receiver that guarantees excellent baselines. The rotating horn was switched between two fixed positions with a frequency of 1~Hz and a beam throw of 120\arcsec. The beam efficiency was $\sim$\,0.53 for a beam size of 40\arcsec. We performed hourly pointing checks using sources from the Effelsberg Catalog of pointing and flux density calibration \citep{ott94}. For the flux calibration we used continuum cross scans of NGC~7027 with a flux density of 5.9~Jy \citep[taken from][]{baars77,mau87}.

We reduced and analyzed the data using the GILDAS\footnote{http://www.iram.fr/IRAMFR/GILDAS} CLASS package. If applicable all spectra have been averaged. To all spectra a baseline has been fitted and was subtracted. In addition, each subscan was individually corrected for the elevation dependency of the telescope gain.

The intensity errors $\Delta$I (see Table\,\ref{tab:properties} for the results) were determined following the procedure from
\citet{bert07}. The geometric average of the line error $\Delta$I$_{\rm L}$\,=\,$\sigma\,v_{\rm res}$\,$\sqrt{N_{\rm L}}$ and the baseline error $\Delta$I$_{\rm B}$\,=\,$\sigma\,v_{\rm res}$\,N$_{\rm L}$/$\sqrt{N_{\rm B}}$ were taken into account. $\sigma$ is the rms noise in Jy, $v_{\rm res}$ the spectral resolution in km\,s$^{\rm -1}$, $N_{\rm L}$ is the number of channels over which the line spreads and $N_{\rm B}$ is the number of channels used for fitting a polynomial to the baseline.

%__________________________________________________________________

\begin{figure*}
\centering
\includegraphics[scale=0.6]{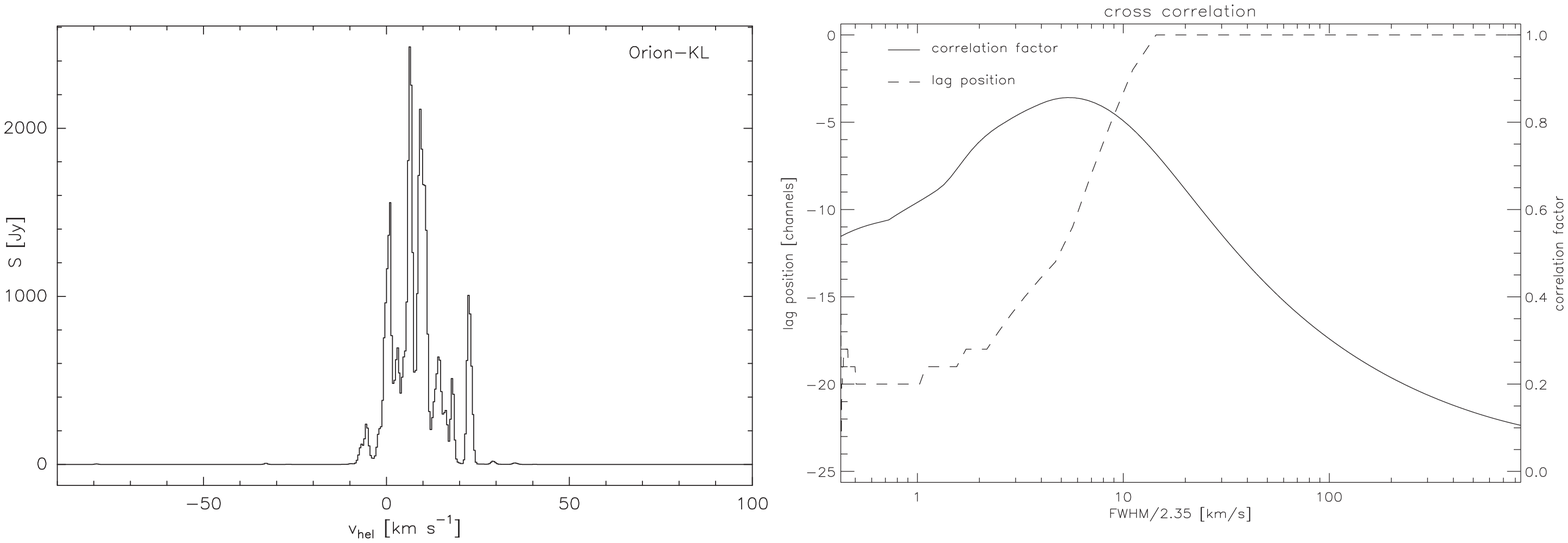}
\caption{\textbf{Left:} Observed spectrum of the galactic known maser source Orion$-$KL at 22~GHz. The channel width is 0.41~km\,s$^{\rm -1}$ \textbf{per channel} and the rms has a value of 0.04~Jy \textbf{per channel}. \textbf{Right:} Results of the cross correlation between the 22~GHz spectrum of Orion$-$KL and Gaussian profiles of different line widths.}
\label{fig:spec+crosscorr_Orion-KL}
\end{figure*}

\begin{figure*}
\centering
\includegraphics[scale=0.6]{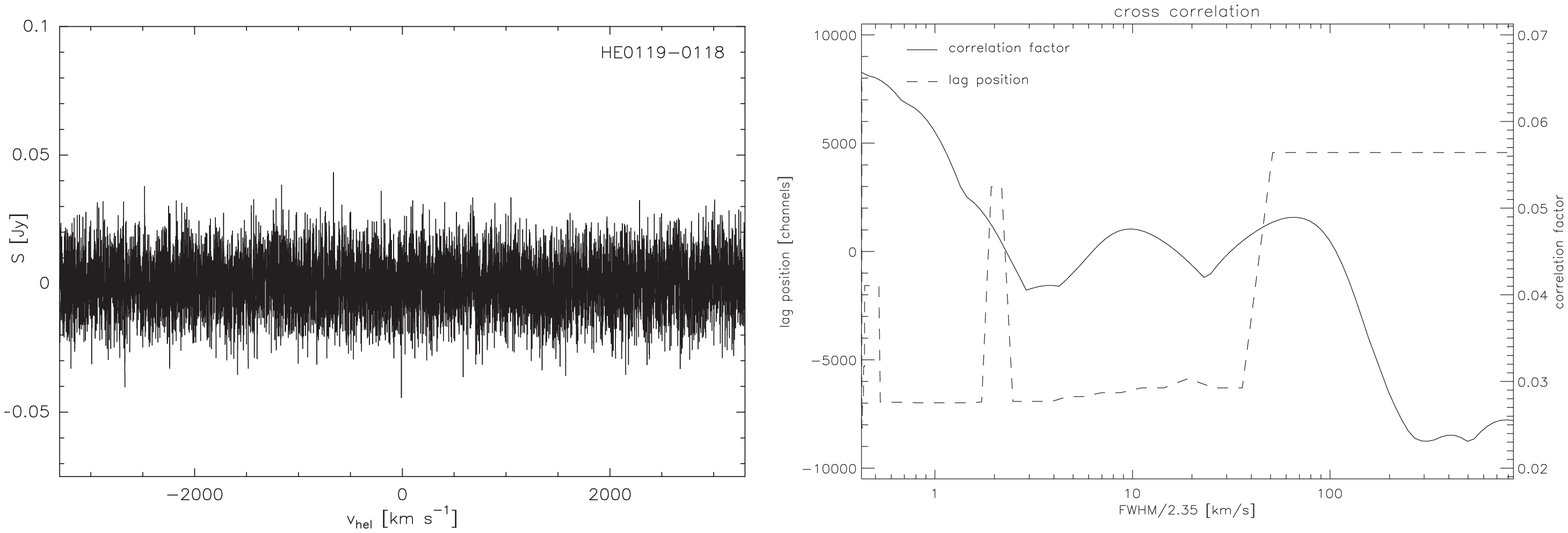}
\caption{\textbf{Left:} Observed spectrum of the nearby QSO host galaxy HE\,0119-0118 at 22~GHz with a channel width of $\sim$\,0.43~km\,s$^{\rm -1}$ \textbf{per channel} and an rms of 0.013~Jy \textbf{per channel}. \textbf{Right:} Results of the cross correlation between the 22~GHz spectrum of HE\,0119$-$0118 and Gaussian profiles of different line widths.}
\label{fig:spec+crosscorr_HE0119}
\end{figure*}

\section{Results$/$Discussion} \label{section:maser_results}

%    We found no H$_{\rm 2}$O maser in our target sample.
%
The observation of 17 objects in our sample of low-luminosity QSO host galaxies yielded no water megamaser detections at individual 3$\sigma$ levels of 27 up to 60~mJy (Table\,1).\\
\indent
To prove that our observational method worked, we observed the known galactic maser sources Orion$-$KL and W3(OH) (see Table\,\ref{tab:properties}). The observed spectrum for Orion$-$KL is shown in Fig.\,\ref{fig:spec+crosscorr_Orion-KL}.\\
\indent
Following the procedure of \citet{gold08} we assumed that all the host galaxies do exhibit H$_{\rm 2}$O megamaser emission at a weak level. To allow for different maser emission velocities in different sources, we adopt the following procedure. At first we select a section of the spectrum around the velocity of CO emission for each host. The offset between the central velocity and the individual components of the megamaser emission can easily be about several hundreds of km\,s$^{\rm -1}$ \citep[for examples see e.g.,][]{kon06b,green07,bra08}. Therefore we used the whole spectral bandwidth as the section to by analyzed. Each spectrum is then cross-correlated with a Gaussian whose width represents the expected line width of the maser emission. In the case of a well-known line width of the megamaser emission each spectrum would then be shifted by the offset between the central channel and the channel of maximum cross-correlation amplitude. To achieve the adjustment to the same velocity offset for all sources the reference channel of the central velocity (formerly defined by the heliocentric velocity of the observed galaxy) in each spectrum has been artificially set to zero. Following this procedure, the shifted data for all observed sources were added together. The described procedure will create a line feature resembling the correlation template (a Gaussian signal in this case) from the constructive alignment of purely random noise. But since the line width of water megamasers can differ between few km\,s$^{\rm -1}$ \citep[NGC~520: FWHM\,=\,1.1~km\,s$^{\rm -1}$,][]{cas08} over several tens \citep[NGC~2989: FWZI\,=\,30~km\,s$^{\rm -1}$,][]{bra08} up to even some hundred km\,s$^{\rm -1}$ \citep[UGC~3193: FWZI\,$\sim$\,350~km\,s$^{\rm -1}$,][]{bra08} we decided to do a study of the cross-correlation parameters first.\\
\indent
Each spectrum was correlated with several Gaussian profiles. The line widths of the Gaussians range from the value of the spectral resolution for each individual spectrum up to 2000~km\,s$^{\rm -1}$. From the cross-correlation we derived the correlation factor and the lag between the peak positions of the Gaussian and the spectrum itself. In order to compare the so gained values to sources known to exhibit maser emission we also performed the correlation on the two galactic objects Orion$-$KL and W3(OH). For a plot of the results on Orion$-$KL see Fig.\,\ref{fig:spec+crosscorr_Orion-KL}. As a representative member of the QSO hosts see the plot on the results of the cross-correlation on HE\,0119$-$0118 (Fig.\,\ref{fig:spec+crosscorr_HE0119}).\\
\noindent
In Fig.\,\ref{fig:spec+crosscorr_Orion-KL} the correlation factor clearly peaks at a FWHM/2.35 of 5.35~km\,s$^{\rm -1}$ which
corresponds to a FWHM of 12.60~km\,s$^{\rm -1}$ of the Gaussian profile. The lag is the number of channels by which the velocity corresponding to the central reference channel has to be shifted left or right in order to match the reference channel to the channel of maximum cross-correlation amplitude. As the correlation coefficient for Orion$-$KL has a maximum (0.856) at a lag of -18 channels the Gaussian has to be shifted by 18 channels to the right hand side of the velocity reference channel in the spectrum to get the maximum agreement between Gaussian profile and spectrum.\\
\indent
Correlation factor and lag position in Fig.\,\ref{fig:spec+crosscorr_HE0119} however show a different behavior. The curve for the correlation factor peaks at three different FWHM. Furthermore the factor ranges between factors of only 0.02 and 0.07 indicating the absence of a correlation. Beyond that the lag position oscillates between shifts of -7000 and +5000 channels. The same behavior can be seen in the plots for the cross-correlation of the other QSO host galaxies.\\
\indent
Although the procedure of the cross-correlation described above could be a measure to distinguish between detections and non-detections, it has some caveats. First of all, the analysis of data with this method may only be appropriate for observations with a good signal-to-noise ratio. Secondly, the FWHM of the line should be known up to some extent. Furthermore one should always keep an eye on the correlation factor and the development of the lag. Too small correlation factors are not trustable in terms of the
confirmation of a possible detection. If the spread of the lag oscillation grows too wide (see e.g., Fig.\,\ref{fig:spec+crosscorr_Orion-KL}) the central line position is not stable, hence indicating either several line components or signals from random noise or radio frequency interference.

\begin{figure*}
\centering
\includegraphics[scale=0.5,angle=-90]{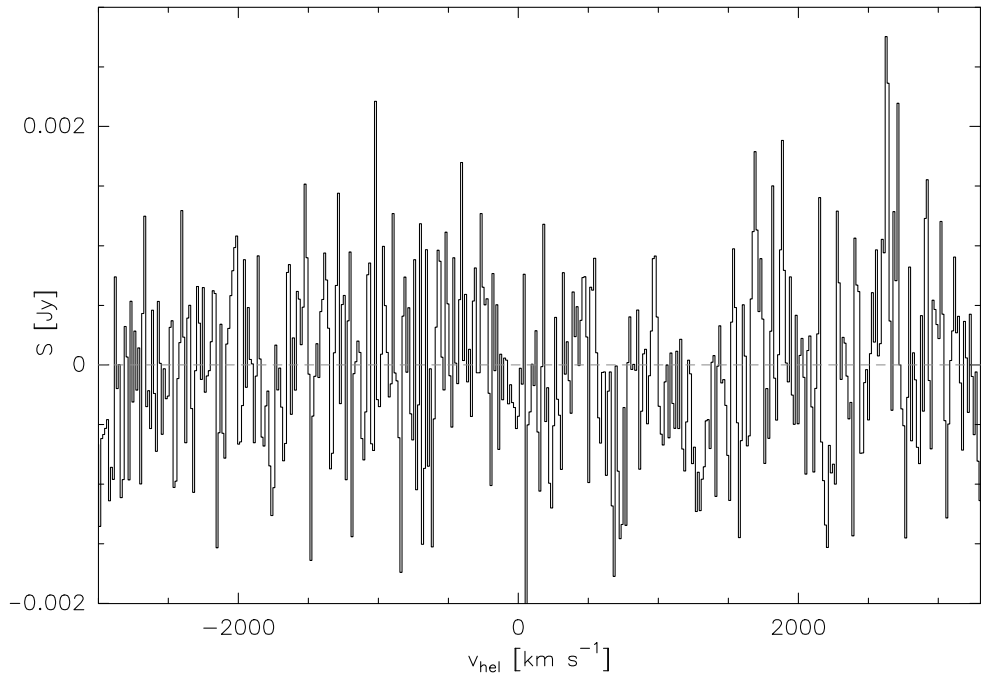}
\caption{Smoothed FFTS spectrum of the resampled and averaged observations of all nearby QSO host galaxies. Before averaging, the spectra were resampled to the same velocity resolution of 0.4367~km\,s$^{\rm -1}$ \textbf{per channel} and the central velocities were shifted to zero. The channel spacing of the displayed spectrum is $\sim$\,14~km\,s$^{\rm -1}$ \textbf{per channel} and the rms is 0.001~Jy \textbf{per channel}. For the averaging process the spectrum of each observed source was weighted equally.}
\label{fig:spec_resampled}
\end{figure*}

\subsection{Sensitivity}

\begin{figure*}[t]
\centering
\includegraphics[scale=0.5,angle=-90]{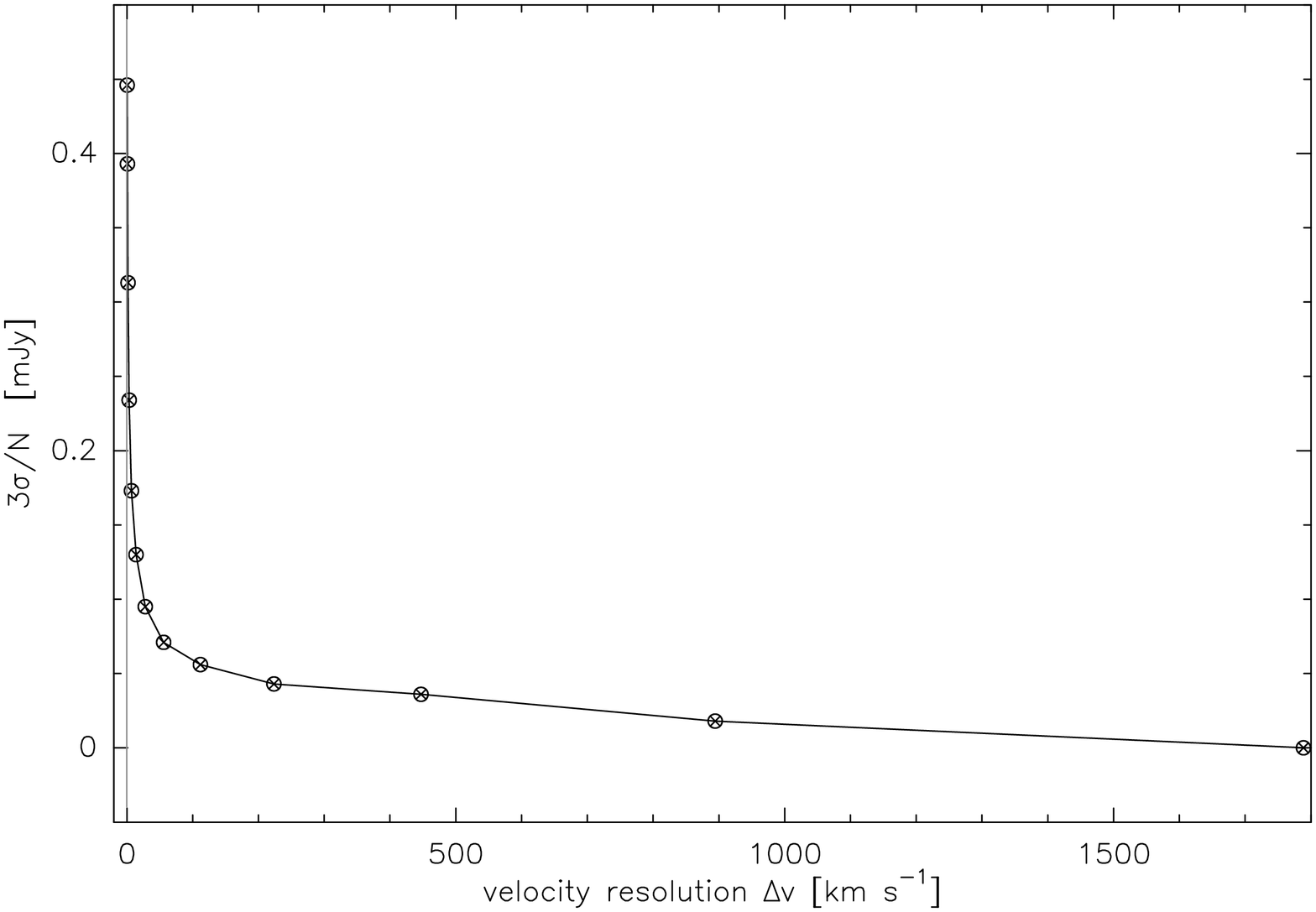}
\caption{3\,$\sigma$ over N (number of observed sources in the sample) as a function of spatial resolution as a mean upper limit of emission in each galaxy.}
\label{fig:3sigma}
\end{figure*}

As \citet{ben09} already pointed out, surveys for H$_{\rm 2}$O megamaser emission put high requirements on the conditions during the observations. Since the water emission is usually very weak ($\ll$\,1~Jy), a high sensitivity is required. The widths of maser lines can be of the order of km\,s$^{\rm -1}$, or even below ($\leq$\,1~km\,s$^{\rm -1}$), but also of the order of hundreds of km\,s$^{\rm -1}$ \citep[e.g., UGC~3193;][]{bra08}. This factor makes a high spectral resolution during the observations a prerequisite. Since the offset between the central velocity of the observed object and the individual components of the megamaser emission can easily be separated by about several hundreds of km\,s$^{\rm -1}$, the bandwidth is another delimiting factor which plays a big role in the observations of water megamaser emission lines. With a velocity coverage of $\sim$\,7200~km\,s$^{\rm -1}$ we provide enough band width (effective 420~MHz with the FFTS) to cover a large velocity range in order not to miss red- or blue-shifted maser components.\\
\indent
Since we don't see any trend for one prominent FWHM for the Gaussian line profiles in the cross-correlation plots, we do not execute the lag shifts but rather add all spectra together after resampling them all to the same spectral resolution. This assumes that the emission occurs at the same velocity offset with respect to the velocity of the central channel in any given spectrum. The resulting resampled and averaged spectrum is shown in Fig.\,\ref{fig:spec_resampled}.\\
\indent
The limits on the maser intensity were determined under the assumption of a FWZI (full width at zero intensity) of
350~km\,s$^{\rm -1}$, which is one of the values for the broadest water maser lines \citep[UGC~3193;][]{bra08}. For our sample sources an average rms noise value of 12~mJy (for $\sim$\,0.4~km\,s$^{\rm -1}$ wide channels) was obtained. In comparison to this, we looked at the values for samples in the literature where water megamaser have been detected. \citet{bra96} state a sensitivity of 60~mJy in $\sim$\,1~km\,s$^{\rm -1}$ channels. \citet{green03a} and \citet{kon06a} observed at a rms of 14~mJy in 1.3~km\,s$^{\rm -1}$ channels. In 2006, \citeauthor{kon06b} published a second paper stating an average rms noise in a 24.4~kHz channel of 4.6~mJy. \citet{bra08} and \citet{cas08} published their observations with rms noise levels of 6~mJy per 0.33~km\,s$^{\rm -1}$ channel and an average rms of 14.33~mJy. In this timely order of observations throughout the years there is the clear tendency that the sensitivity increases with time significantly. With our rms value of 12~mJy we are well within the range where detections in theory should be possible.\\
\indent
In order to get a handle on the upper limit of the emission from the observed frequencies at 22~GHz, we determined the
3\,$\sigma$-over-N value for the resampled spectrum for 13 different velocity resolutions (Fig.\,\ref{fig:3sigma}). $\sigma$ is the channel-to-channel rms noise and N is the number of host galaxies observed in our sample. Fig.\,\ref{fig:3sigma} shows that with growing channel width the 3\,$\sigma$/N value goes down, which means that by accumulating signal through the binning of several channels the noise decreases. Since in this paper we report only non-detections of H$_{\rm 2}$O megamaser emission, we use this plot as a measure to get an estimation of a mean upper limit of the emission in each sample galaxy.

\begin{figure*}
\centering
\includegraphics[scale=0.5,angle=-90]{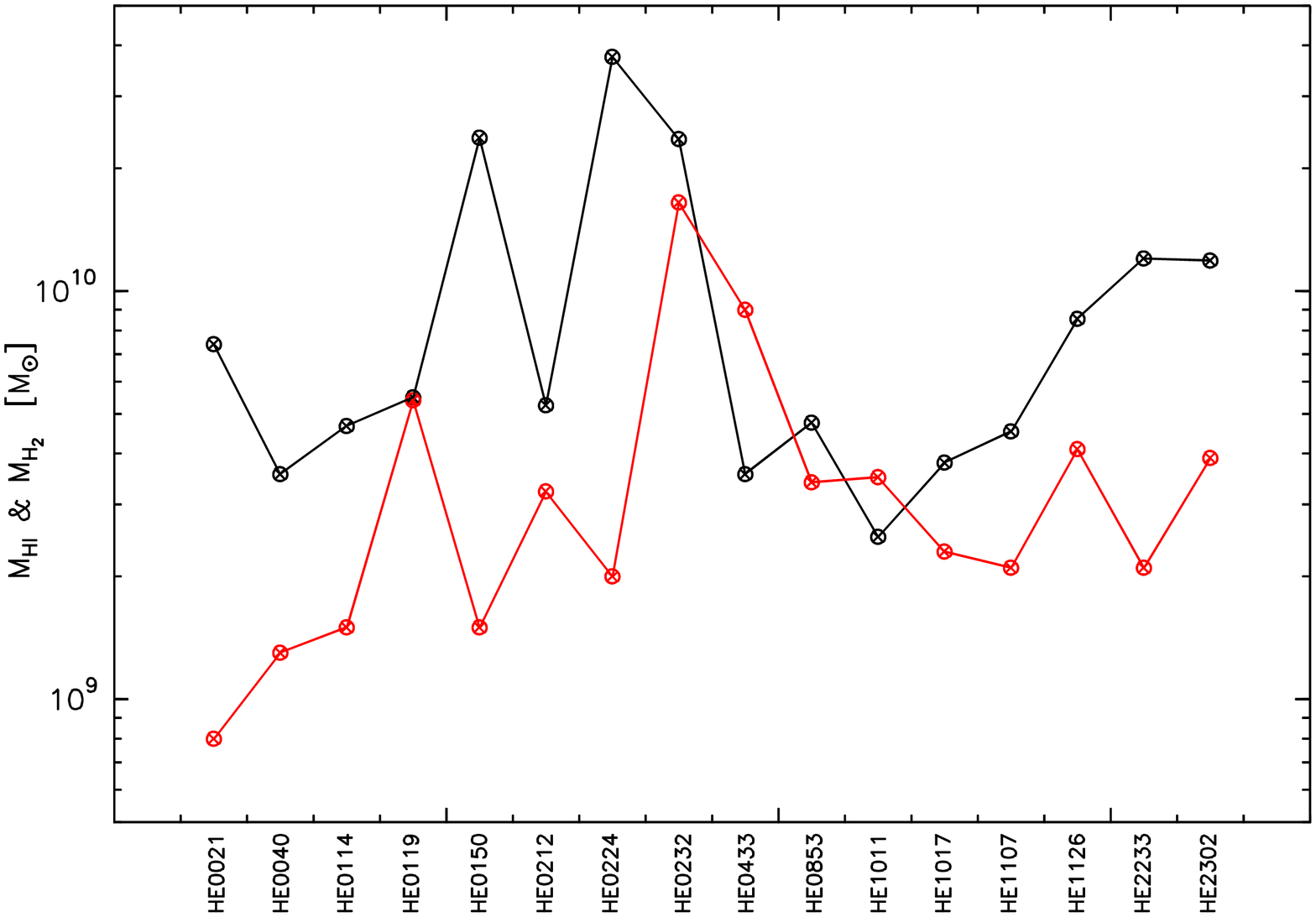}
\caption{Comparison of the atomic and molecular gas masses of the sources searched for water maser emission. Values for H{\sc{I}} masses are depicted by black circles, red circles represent the H$_{\rm 2}$ masses derived from CO observations.}
\label{fig:source_HI_H2}
\end{figure*}

\begin{figure*}
\centering
\includegraphics[scale=0.5]{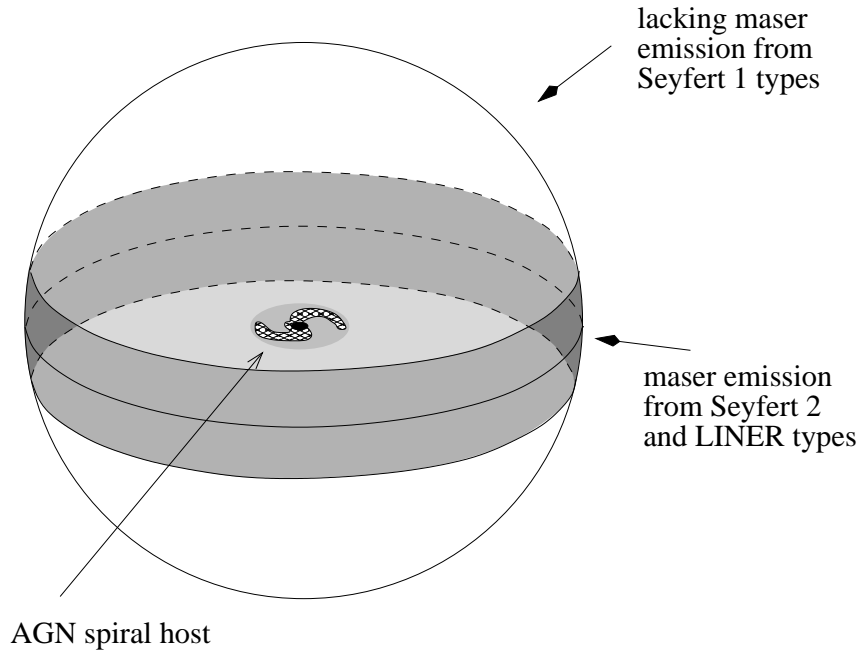}
\caption{A simple depiction of the unified scheme in context of water megamaser emission. Sy~2 galaxies are the host
galaxies with the most water maser detections. A possible explanation in terms of the unified scheme is that the only difference between Sy~1 galaxies and Sy~2s is the viewing angle of the observer. Sy~2 galaxies are seen more edge-on whereas Sy~1s are seen more pole-on. The angle under which the maser emission is detectable is only $\pm$\,6\degr\ from the equatorial plane (region shaded in grey). This explains the lack of megamaser detections in Sy~1 galaxies.}
\label{fig:cartoon}
\end{figure*}

\subsection{Host galaxy properties}

16 of the 17 observed sources were detected in molecular gas emission \cite[CO(2--1) \& CO(1--0),][]{bert07}. In addition, 5 of the CO detected sources which were searched for H$_{\rm 2}$O megamaser emission have been detected previously in 21~cm H{\sc{I}} line emission \citep{koenig09}. Fig.\,\ref{fig:source_HI_H2} shows the neutral atomic H{\sc{I}} gas mass M$_{\rm H{\sc{I}}}$ \citep[values from][]{koenig09} and the molecular gas mass M$_{\rm H_2}$ derived from CO observations \citep[values from][]{bert07}. It is surprising to see that for 14 of the 16 nearby QSO hosts with the highest luminosities in the IR, the H{\sc{I}} mass is larger than the molecular gas mass. Nonetheless, this seems to be only a 'local' trend in this sample. To achieve statistical significance on this subject the sample considered is too small.

\subsection{Morphology}

78 maser galaxies have been detected so far \citep{ben09}. The water megamaser observed at the highest redshift is
MG~J0414+0534 at z\,=\,2.639 \citep{imp08}. The biggest part in the population of masing galaxies is made up by Seyfert types
(78\%). 11\% are LINERs (low ionization nuclear emission regions) and 7\% are H{\sc{II}} galaxies. The smallest percentage of
galaxies showing water megamaser signatures are starburst galaxies (3\%) and NLRGs (narrow-line radio galaxies). Although Seyfert galaxies present the majority of maser galaxies only 3\% are Sy~1s, whereas Sy~2s are the dominant type (88\%). Even though past surveys were more focused on Seyfert type galaxies than for example on Fanaroff \& Riley class sources \citep[FR,][]{far74} a similar dichotomy is observed for FRI and FRII galaxies. 50 FRI sources \citep[average redshift: 0.040,][]{hen98} have been searched for water megamaser emission, yielding no detection. For FRII sources a smaller sample of 3 sources 
\citep[average redshift: 0.056,][]{tar03} was observed resulting in one detection of H$_{\rm 2}$O lines. 
\citet{ben09} find that most known water megamaser galaxies are classified as spirals (84\%). The remaining 16\% are made up by S0 galaxies (7\%), elliptical galaxies (1\%) and irregular or peculiar galaxies (8\%). Only NGC~1052, classified as Sy~2
galaxy or LINER has an elliptical morphology \citep{ben09}. They argue that the mechanism fueling the nuclear activity is one
important property separating spirals (morphological type of most Seyferts) from early type galaxies (morphological type of most QSOs). Quasars of both classes have been searched for water megamaser emission: H$_{\rm 2}$O in type 1 quasars was searched for in only a few targets (17 sources, average redshift: 0.049, this work; 1 source, redshift: 2.639, Impellizzeri et al. 2008), while much larger samples of type 2 quasars were investigated (47 sources, redshift: 0.3\,$<$\,z\,$<$\,0.8, Barvainis \& Antonucci 2005; 274 sources, redshift: 0.3\,$<$\,z\,$<$\,0.8, Bennert et al. 2009). As a result of these surveys H$_{\rm 2}$O megamaser emission was already detected not only in a type 2 quasar \citep{bar05}, but also in a type 1 quasar \citep{imp08}, which, furthermore, is the megamaser with the highest redshift. Compared to the number of Seyfert galaxies searched for H$_{\rm 2}$O emission these numbers are very small: More than 600 Seyfert galaxies have been observed in the 22~GHz water line \citep[e.g.,][]{bra96,green02,bra04,kon06a,zhang06,bra08}.\\
\indent
\citet{bra97} found that the absence of detections in Seyfert~1 galaxies indicates either that these galaxies do not have molecular gas with appropriate conditions to mase, or that the masers in these galaxies are beamed away from our line of sight. The latter is in good agreement with the findings of \citet{bra08}, who state that masers, specifically those in AGN accretion disks are beamed in the plane of the disk. \citet{miyo95} show with VLBA observations of NGC~4258 that the 
H$_{\rm 2}$O emission arises from a thin edge-on disk only a fraction of a parsec away from the supermassive black hole at the nucleus. The unified scheme of AGN \citep{law82,anto93} says that Sy~1 nuclei are hidden within Sy~2 galaxies behind an obscuring dusty molecular thick disk, or torus, indicating that the two different types of Seyfert galaxies only differ in terms of the viewing angle: Sy~1s are seen more pole-on while Sy~2s are seen more edge-on. If one takes the detection rate of H$_{\rm 2}$O megamasers in all AGN (10\%) into account, the probability to find no maser emission is at 90\%. We tried to determine the viewing angle under which the probability to find megamaser emission is 10\%.
\begin{equation}
 P(\vartheta) = 2\,\cdot\,\frac{2\pi\,(1\,-\,\cos\,\vartheta)}{4\,\pi} = 1\,-\,\cos\,\vartheta
\label{equation:probability}
\end{equation}
For the angle $\vartheta$ a value of 0\degr\ describes an edge-on view onto the the galaxy, 90\degr\ represents the face-on view. Putting the values into equation (\ref{equation:probability}) results in an angle of 6\degr. This means that the maser emission most probably will be detected if the line of sight to the observer falls within an angular distance of $\pm$\,6\degr\ from the equatorial plane. For that see also Fig.\,\ref{fig:cartoon}.\\
\indent
Our sample consists of eight elliptical or S0 host galaxies, six spirals and three galaxies that seem to be in a merger/interaction phase, which unfortunately decreases the, a priori already small, chance to find a water maser in a sample composed of Sy~1$-$1.5, even more. Taking the statement of \citet{ben09} about the morphology of the galaxies known to show maser emission into account the chance of finding one maser from an elliptical host galaxy is very slim. Statistically, one out of 78 detected water megamaser sources shows elliptical morphology, which means it would take $\sim$\,1900 observed galaxies to find a second elliptical host galaxy with water megamaser emission. In addition, the continuum fluxes are very low for our sample sources. A strong continuum could amplify the 22~GHz H$_{\rm 2}$O maser emission. Jets are not known for any of the sample galaxies.

\subsection{Black hole mass}

One prominent difference between Seyfert galaxies and QSOs, besides the brightness on the absolute magnitude scale, lies in the mass of their nuclear engines. Seyferts have black hole masses between 10$^{\rm 6}$ and few 10$^{\rm 7}$~M$_{\sun}$
\citep[e.g.,][]{her99,hen02,kuo11}. Quasars, on the other hand, can reach masses of the central black hole up to 10$^{\rm 9}$~M$_{\sun}$ \citep[e.g.,][]{lab06,ves08}. The black hole masses for the sources in our sample range from rather small, at least for QSOs, 1.07\,$\times$\,10$^{\rm 7}$~M$_{\sun}$ \citep[HE\,1011$-$0403,][]{wang01} up to large 3.47\,$\times$\,10$^{\rm 8}$~M$_{\sun}$ \citep[HE2302$-$0857,][]{oneill05}. The large black hole masses could possibly imply that the conditions in the vicinity of the nuclear black hole are not stable enough to cause strong megamaser emission \citep{tar07,ben09}.

\section{Conclusions/Summary} \label{section:maser_conclusions}

We present the results on our search for H$_{\rm 2}$O water megamasers in 17 nearby low-luminosity QSO host galaxies. In none of the target sources we were able to find emission in the 22~GHz maser line. We therefore confirm the results of previous water megamaser surveys stating, that extragalactic water masers are found primarily in Seyfert~2 galaxies and LINERs. We compared the atomic and molecular mass contents of the member galaxies of our sample. For almost all of them the atomic gas content is larger than the molecular one. A sensitivity study shows that the observational setup used obtaining the discussed data was sufficiently suited to detect water megamaser emission. To prove this in practice we observed two known galactic maser sources Orion$-$KL and W3(OH) successfully. We show further that surveys like the one reported in the present work are required in order to enhance the statistics for QSOs and see if there are differences for the megamaser emission between Seyfert and QSO host galaxies.

\section*{Acknowledgments}
  MG-M is supported by the German federal department for education and research (BMBF) under the project numbers: 50OS0502
  \& 50OS0801. We thank the anonymous referee for helpful comments. The results presented in this paper are based on observations with the 100-m telescope of the MPIfR
  (Max-Planck-Institut f\"ur Radioastronomie) at Effelsberg. This research has made use of the NASA/IPAC Extragalactic Database
  (NED) which is operated by the Jet Propulsion Laboratory, California Institute of Technology, under contract with the National
  Aeronautics and Space Administration.

\bibliographystyle{mn2e}
\bibliography{maser_mnras}

%__________________________________________________________________________________________________________________________________
%__________________________________________________________________________________________________________________________________

%\appendix
%
%\section[]{Large gaps in L\lowercase{y}${\balpha}$ forests\\* due to fluctuations in line distribution}

%\bsp

\label{lastpage}

\end{document}